\documentclass[journal]{IEEEtran}
\usepackage{amsmath,amssymb,latexsym, bm,bbm,mathrsfs}
\usepackage{epsfig}
\usepackage{caption}

\fussy
\newtheorem{theorem}{Theorem}
\newtheorem{definition}[theorem]{Definition}

\newtheorem{lemma}[theorem]{Lemma}

\def\proof{\noindent{\bf Proof}\ \ }
\def\qed{\hfill \vrule height 7pt width 7pt depth 0pt\medskip}

\newcommand{\ds}{\displaystyle}

\newcommand{\ba}{\begin{array}}
\newcommand{\ea}{\end{array}}

\newcommand{\mb}{\boldsymbol}
\newcommand{\be}{\begin{equation}}
\newcommand{\ee}{\end{equation}}
\newcommand{\eps}{\varepsilon}

\newcommand{\mc}{\mathcal}

\newcommand{\ov}{\overline}

\newcommand{\R}{\mathbb{R}}

\newcommand{\summ}{\sum\limits}
\renewcommand{\P}{\mathbb{P}}

\newcommand{\de}{\mathrm{d}}
\newcommand{\prodd}{\prod\limits}

\DeclareMathOperator{\co}{co}

\DeclareMathOperator{\ent}{H}\DeclareMathOperator{\mutinf}{I}

\def\R{\mathbb{R}}

\def\P{\mathbb{P}}

\begin{document}

\title{On the Capacity of Memoryless Finite-State Multiple-Access Channels with Asymmetric State Information at the Encoders}
\author{Giacomo Como \thanks{Giacomo Como is with the Laboratory for Information and Decision Systems, Massachusetts Institute of Technology, 77 Massachusetts Avenue, Cambridge, MA 02139, USA. Email: giacomo@mit.edu.} 
and Serdar Y\"uksel \thanks{S.~Y\"uksel is with the Mathematics and Engineering Program, Department of Mathematics and Statistics, Queen's University, Kingston, Ontario, Canada, K7L 3N6. Email: yuksel@mast.queensu.ca. S.~Y\"uksel's research is supported by the Natural Sciences and Engineering Research Council of Canada (NSERC).}}
       \maketitle

\begin{abstract}
A single-letter characterization is provided for the capacity region of finite-state multiple-access channels, when the channel state process is an independent and identically distributed sequence, the transmitters have access to partial (quantized) state information, and complete channel state information is available at the receiver. The partial channel state information is assumed to be asymmetric at the encoders. As a main contribution, a tight converse coding theorem is presented. 
%A simple proof of achievability is reported as well.
%, which has not been presented earlier to our knowledge and which uses non-standard techniques. %The problem is practically relevant, and admits a single-letter characterization for its solution.
The difficulties associated with the case when the channel state has memory are discussed and connections to decentralized stochastic control theory are presented.

\end{abstract}


\maketitle

\section{Introduction and Literature Review}

Wireless communication channels and Internet type networks are examples of channels where the channel characteristics are time-varying. In wireless channels, the mobility of users and changes in landscape as well as interference may lead to temporal variations in the channel quality. In network applications, user demand and node failure may affect the channel reliability. Such channel variation models may include fast fading and slow fading; in fast fading, the channel state is assumed to be changing for each use of the channel. On the other hand, in slow fading, the channel is assumed to be constant for each finite block length.

In such problems, the channel state can be transmitted to the encoders either explicitly, or through output feedback. Typically the feedback is not perfect, that is the encoder has only partial information regarding the state or the output variables. The present paper studies a particular case, finite-state multiple-access channels (MACs), where partial channel state information (CSI) is provided to the encoders causally. What makes such setup particularly interesting is the fact that the partial CSI available to the two transmitters is in general \emph{asymmetric}, i.e., none of the transmitters' CSI contains the CSI available to the other one. On the other hand, we assume that the receiver has access to perfect state information.

A single-letter characterization of the capacity region is provided for the case of independent and identically distributed (i.i.d.) channel state sequences. As we shall review shortly, results in the literature have already provided achievability results for such problems. The main contribution of this paper consists in providing a tight converse theorem. Our proof involves showing that restricting to encoders using only the quantized CSI on the current state does not cause any loss of optimality with respect to the most general class of admissible encoders using all the quantized CSI causally observed until a given time.

The problem at hand can be thought of as a decentralized stochastic control problem. We shall elaborate on this connection in the concluding section, where we shall also discuss in what our arguments fail when trying to extend them to a proof of the converse theorem for finite-state MACs with memory, and asymmetric CSI at the transmitters.
%
%With this insight, we provide a non-standard argument to show that . We believe this approach will find various applications in multi-terminal information theory. %Later in the paper, we consider also the case when the channel state has memory. In this case, we shall make the observation that we cannot define a finite dimensional sufficient statistic on which the encoders can base their decisions on. The encoders need to use their entire information set for generating channel inputs to obtain an optimal coding scheme.

%The limitations of the proposed techniques to handle the general case where the channel state process is Markovian are discussed, as well as the analogies of this problem with team decision theory \cite{WitsenhausenEqui,Yuksel}.  The latter is concerned with ...... COMPLETE....

Let us now present a brief literature review. Capacity with partial channel state information at the transmitter is related to the problem of coding with unequal side information at the encoder and the decoder. The capacity of memoryless channels, with various cases of state information being available at neither, either or both the transmitter and receiver, has been studied in \cite{Shannon} and \cite{GelfandPinsker}. Reference \cite{TatikondaMitter} develops a stochastic control framework for the computation of the capacity of channels with memory and complete noiseless output feedback via the properties of the directed mutual information. Reference \cite{GoldsmithVaraiya} considers fading channels with perfect channel state information at the transmitter, and shows that with instantaneous and perfect state information, the transmitter can adjust the data rates for each channel state to maximize the average transmission rate. Viswanathan \cite{Viswanathan} relaxes this assumption of perfect instantaneous state information, and studies the capacity of Markovian channels with delayed information. Reference \cite{ChenBerger} studies the capacity of Markov channels with perfect causal state information. The capacity of Markovian, finite-state channels with quantized state information available at the transmitter is studied in \cite{YukTatISIT07}. %This section extends the findings of \cite{YukTatISIT07}.

\begin{figure}
\centering
\epsfig{figure=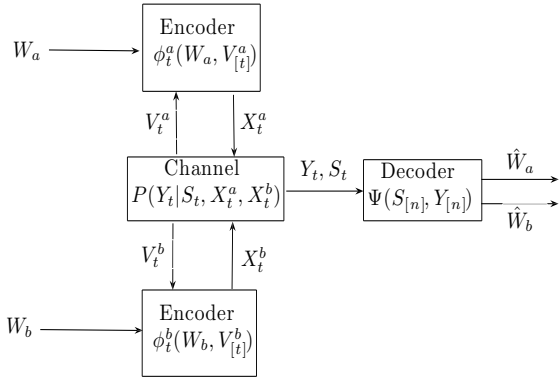,height=6cm,width=8cm}
\caption{Finite-state multiple-access channel with asymmetric partial state information at the transmitters. \label{LL}}
\end{figure}

The works most closely related to ours are  \cite{DasNarayan} and \cite{Jafar}. In \cite{DasNarayan}, the capacity of general finite-state MACs with different levels of causal CSI at the transmitters is characterized in terms of multi-letter formulas. Moreover, single-letter characterizations are provided for the capacity of finite-state MACs when the decoder has perfect CSI and the encoders are restricted to use only a finite window of, possibly limited, CSI; the capacity region without any such restriction is recovered in the limit of large window size. Reference \cite{Jafar} develops a general framework for approximating, and possibly characterizing, the capacity of channels with causal, and non-causal CSI: in particular, Theorem 4 therein provides a single-letter characterization of the capacity region of a MAC with independent CSI at the transmitters. With respect to \cite{DasNarayan,Jafar}, the present paper considers the somewhat simpler case of a MAC with i.i.d.~state, where the encoders have causal, asymmetric, partial CSI, which is obtained through fixed quantizers acting componentwise. In contrast to \cite{DasNarayan}, a single-letter expression for the capacity region is obtained in this case without any finite window restriction on the CSI available to the transmitters, while, differently from \cite{Jafar}, the CSI available to the transmitters is not assumed to be independent. Recent related work also includes \cite{Permuter}, providing an infinite-dimensional characterization for the capacity region for Multiple Access Channels with feedback, and \cite{CemalSteinberg}, studying the case of MAC channels where the encoders have access to coded non-causal state information.
%Authors in  \cite{CemalSteinberg} also considered outer and inner bounds on the capacity region when the information at the encoders is asymmetric, that is not necessarily degraded.

The rest of the paper is organized as follows. In Section \ref{sectmainresults} a formal statement of the problem and the main results are presented, consisting in a single-letter characterization of the capacity region of finite-state MACs with i.i.d. state.
Section \ref{sectachievability} contains the proof of achievability of the capacity region, while Section \ref{sectconverse} presents a proof of the converse coding theorem. Finally, in Section \ref{sectconclusions}, we discuss the issues arising when trying to generalize our arguments to the memory case, and present some final remarks on the connections of this problem with the decentralized stochastic control literature.

\section{Capacity of i.i.d.~finite-state MAC with asymmetric partial CSI}
\label{sectmainresults}

In the following, we shall present some notation, before formally stating the problem.
For a vector $v$, and a positive integer $i$, $v_i$ will denote the $i$-th entry of $v$,
while $v_{[i]}=(v_1,\ldots,v_{i})$ will denote the vector of the first $i$ entries of $v$.
Following a common convention,  %for a random variable,
capital letters will be used to denote random variables (r.v.s), and small letters denote particular realizations.
We shall use the standard notation $\ent(\,\cdot\,)$, and $\mutinf(\,\cdot\,;\,\cdot\,)$ (respectively $\ent(\,\cdot\,|\,\cdot\,)$, and $\mutinf(\,\cdot\,;\,\cdot\,|\,\cdot\,)$)
for the (conditional) entropy and mutual information of r.v.s.
With a slight abuse of notation, for $0\le x\le1$, we shall write $\ent(x)$ for the entropy of $x$.
For a finite set $\mc A$, $\mc P(\mc A)$ will denote the simplex of probability distributions over $\mc A$.
Finally, for a positive integer $n$, we shall denote by
$\textstyle \mc A^{(n)}:=\bigcup_{0\le s<n}\mc A^s$
the set of $\mc A$-strings of length smaller than $n$.
\footnote{This includes the empty string, conventionally assumed to be the only element of $\mc A^0$.}
%In the following, a bold letter ${\bf V}$ will denote an ensemble $(V^1,V^2)$, for general vectors $V^1, V^2$.

We shall consider a finite-state MAC with two transmitters, indexed by $i\in\{a,b\}$, and one receiver.
Transmitter $i$ aims at reliably communicating a message $W_i$, uniformly distributed over some finite message set $\mc W_i$, to the receiver.
The two messages $W_a$ and $W_b$ are assumed to be mutually independent. We shall use the notation $W:=(W_a,W_b)$ for the vector of the two messages.

The channel state process is modeled by a sequence $S=\{S_t:\, t=1,2,\ldots\}$ of independent, identically distributed (i.i.d.) r.v.s, taking values in some finite-state space $\mc S$, and independent from $W$; the probability distribution of any $S_t$ is denoted by $P(\,\cdot\,)\in\mc P(\mc S)$.
%The channel process considered is a finite state, i.i.d. process $\{, t=1,2,\dots\}$ taking values in a finite set ${\cal S}$.
The two encoders have access to causal, partial state information: at each time $t\ge1$, encoder $i$ observes $V^{(i)}_t=q_i(S_t)$,
%the observation variables at the coders $\{V^1, V^2\}$ are generated through
%quantizers  (this represents the )
%such that $P(V^{(i)}_t=\tilde{s}|S_t=s)$ is a function of $s_t$ and $P(.|S_t=s)$ is a probability measure on ${V}^{(i)}$, for every $s \in {\cal S}$,
%$t=0,1,\dots$, $i=1,2$,
where $q_i:\mc S\to\mc V_i$ is a quantizer modeling the imperfection in the state information.
We shall denote by $V_t:=(V^{(a)}_t,V^{(b)}_t)$ the vector of quantized state observations, taking values in $\mc V:=\mc V_a\times\mc V_b$.
%When the channel state is quantized, then the probability measure becomes a Dirac measure on the quantized value.
The channel input of encoder $i$ at time $t$,  $X^{(i)}_t$, takes values in a finite set $\mc X_i$, and is assumed to be a function of the locally available information $(W_i,V^{(i)}_{[t]})$. The symbol $X_t=(X^{(a)}_t,X^{(b)}_t)$ will be used for the vector of the two channel inputs at time $t$, taking values in
$\mc X:=\mc X_a\times\mc X_b$.
The channel output at time $t$, $Y_t$, takes values in a finite set ${\cal Y}$; its conditional distribution satisfies
\be\!\P(Y_t\!=\!y|W\!=\!w,X_{[t]}\!=\!x_{[t]},S_{[t]}\!=\!s_{[t]})\!=\!P(y_t|s_t,x_t)\label{channel}\ee
where, for any $s\in\mc S$, and $x\in\mc X$, $P(\,\cdot\,|s,x)\in\mc P(\mc Y)$ is an output probability distribution.
Finally, the decoder is assumed to have access to perfect causal state information (which may be known causally or non-causally);
the estimated message pair will be denoted by $\hat W=(\hat W_a,\hat W_b)$.

We now present the class of transmission systems.
\begin{definition}
For a rate pair $R=(R_a,R_b)$, a block-length $n\ge1$, and a target error probability $\eps\ge0$, an $(R,n,\epsilon)$-coding scheme consists of two sequences of functions
$$\{\phi^{(i)}_t:\mc W_i\times\mc V_i^t\to\mc X_i\}_{1\le t \le n}\,,$$
and a decoding function
$$\psi:\mc S^n\times\mc Y^n\to\mc W_a\times\mc W_b\,,$$
such that, for $i\in\{a,b\}$, $1\le t\le n$:
\begin{itemize}
\item $|\mc W_i|\ge2^{R_in}$;%\\[2pt]
\item $X^{(i)}_t=\phi^{(i)}_t(W_i,V^{(i)}_{[t]})$;%\\[2pt]
%The team coding policy is a sequence of functions $\{f^{(i)}_t, k \geq 1 \}$ of $(W_i,V^{(i)}_{[t]},X^{(i)}_{[t-1]})$ (with $V^{(i)}_0=\tilde{s}_0$) and with range spaces $\mc X_i$. The channel code output at time
%$t$ is $f^{(i)}_t[w](\tilde{s}_{[t]})$.
\item $\hat W:=\psi(S_{[n]},Y_{[n]})$; %\\[2pt]
\item $\P(\hat W\ne W)\le\eps$.
%
%A channel decoder function $g$ maps $({\cal Y}_{[N]},{\cal S}_{[N]}) \to {\cal W}^1 \times {\cal W}^2$, such that the (average) probability of error satisfies $P_e \leq \epsilon$, with $P_e$:
%\begin{eqnarray}
%&& {1 \over |{\cal W}^1 \times {\cal W}^2|} \sum_{(w^1,w^2) \in {\cal W}^1 \times {\cal W}^2} \nonumber \\
%&& \quad \quad Pr\bigg(g(Y_{[n]},S_{[n]}) \neq (W_a,W_b) \nonumber \\
%&& \quad \quad \quad \quad \bigg| (W_a,W_b)=(w^1,w^2)\bigg) \leq \epsilon,
%\end{eqnarray}
%and $|{\cal W}^{(i)}|=M_i$, i=1,2.
\end{itemize}
%We shall denote by $\phi_t:\mc W\times\mc V^t\to \mc X$,
%$$\phi_t(w,v_{[t]}):=(\phi_t^{(a)}(w_a,v^{(a)}_{[t]}),\phi_t^{(b)}(w_b,v^{(b)}_{[t]}))\,,$$
%the joint encoder.
\end{definition}

We now proceed with the characterization of the capacity region.
\begin{definition}
A rate pair $R=(R_a,R_b)$ is achievable if, for all $\eps>0$, there exists, for some $n\ge1$, an $(R,n,\eps)$-coding scheme.
%$R^{(i)}$ is an $\epsilon-$achievable rate if for every $\delta > 0$, there exists, for all sufficiently large $N$, an $(\mc W_1,\mc W_2,n,\epsilon)$ code such that ${1 \over N}\log_2(M_i) \geq R^{(i)} -\delta$. $R_i$ is achievable
%if it is $\epsilon-$achievable for all $\epsilon > 0$.
The capacity region of the finite-state MAC is the closure of the set of all achievable rate pairs.
\end{definition}

We now introduce what we call {\em memoryless stationary team policies} and their associated rate regions.
%(MAYBE WE DON'T WANT TO CALL IT THIS WAY, BUT ONLY USE THIS TERMINOLOGY IN THE CONCLUSION)
\begin{definition}
A memoryless stationary team policy is a family
\be\label{policy} \pi=\left\{\pi_i(\,\cdot\,|v_i)\in\mc P(\mc X_i)|\,i\in\{a,b\},\,v_i\in\mc V_i\right\}\ee
 of probability distributions on the two channel input sets conditioned on the quantized observation of each transmitter.
For every memoryless stationary team policy $\pi$, $\mc R(\pi)$ will denote the region of all rate pairs $R=(R_a,R_b)$ satisfying
\be\label{region}
\ba{rcccl}
0&\le& R_a &<& \mutinf(X_a;Y| X_b,S) \\
0&\le& R_b &<& \mutinf(X_b;Y| X_a,S) \\
0&\le& R_a+R_b&<& \mutinf(X;Y| S)\,,\ea
\ee
where $S$, $X=(X_a,X_b)$, and $Y$, are r.v.s taking values in $\mc S$, $\mc X$, and $\mc Y$, respectively, and whose joint probability distribution
$$\nu(s,x,y):=P(S=s,X=x,Y=y)$$
factorizes as
\be\label{factorization}\nu(s,x,y)= P(s)\pi_a(x_a|q_a(s))\pi_b(x_b|q_b(s))P(y|s,x)\,.\ee
\end{definition}

%$\mc R(\mu_a,\mu_b)$ the set of rate pairs
We can now state the main result of the paper.
\begin{theorem}\label{theomain}
The achievable rate region is given by
$$\ov{\co}\left(\bigcup\nolimits_{\pi}\mc R(\pi)\right)\,,$$
the closure of the convex hull of the rate regions associated to all possible memoryless stationary team policies $\pi$ as in (\ref{policy}).
\end{theorem}
In Section \ref{sectachievability} we shall prove the direct part of Theorem \ref{theomain}, namely that every rate pair $R\in\ov{\co}\left(\cup_{\pi}\mc R(\pi)\right)$
is achievable. In Section \ref{sectconverse} we shall prove the converse part, i.e.~that no rate pair $R\in\R_+^2\setminus\ov{\co}\left(\cup_{\pi}\mc R(\pi)\right)$
is achievable.

\section{Achievability of the capacity region}
\label{sectachievability}
%IN ORDER TO MAKE ELZA HAPPY WE MAY WONNA SHORTEN THIS SECTION, AND POSSIBLY MERGE IT TO THE FOLLOWING

The result on achievability is known, and follows, e.g., from \cite{DasNarayan}. For convenience, we briefly sketch a different approach, as suggested at the beginning of \cite[Sect.~VI]{Jafar}. The main idea consists in considering an equivalent MAC having as input mappings form the CSI information available at the transmitters to the original MAC's input. Specifically, one considers an equivalent memoryless MAC having output space $\mc Z:=\mc S\times\mc Y$ coinciding with the product of the state and output space of the original MAC, input spaces $\mc U_i:=\{u_i:\mc V_i\to\mc {\cal X}_i\}$, for $i\in\{a,b\}$, and transition probabilities
$$Q(z|u_a,u_b):=P(s)P(y|u_a(q_a(s)),u_b(q_b(s)))\,,$$
where $z=(s,y)$.
%A coding scheme for such a MAC consists of a pair of encoders $f^{(i)}:\mc W_i\to\mc U_i^n$, $i\in\{a,b\}$, %and $f^2:\mc W_b\to{\mc {\cal U}^2}^n$,
%and a decoder $g: \mc Y^n\times\mc S^n\to\mc W_a\times\mc W_b$. To any such coding scheme it is natural to  associate a coding scheme for the original finite-state MAC, by defining the encoders
%$$\phi^{(i)}_t:\mc W_i\times\mc V_i^t\to\mc X_i\,,\quad\phi^{(i)}_t(w_i,v_{[t]})=[f^{(i)}(w_i)](v^{(i)}_t)$$
%and letting the decoder $\psi:\mc Y^n\times\mc S^n\to\mc W_a\times\mc W_b$
%coincide with $g$.
%%$\pi:{\cal V}^1\to P({\cal U}^1)$, $\pi:{\cal V}^2\to P({\cal U}^2)$,
%%all the rate pairs $(R_1,R_2)$ satisfying
%%\begin{eqnarray}
%%R^1 &\leq& \mutinf(U^1;Y| U^2,S) \nonumber \\
%%R^2 &\leq& \mutinf(U^2;Y| U^2,S) \nonumber \\
%%R^1 + R^2 &\leq& \mutinf(U^1,U^2;Y| S),
%%\end{eqnarray}
%%with respect to the joint distribution
%%$P(S,U^1,U^2,Y) = P(Y|S,U^1,U^2)\pi_1(U^1|V^1)\pi_2(U^2|V^2)P(S)$ are achievable.
%It is not hard to verify that the probability measure induced on the product space $\mc W_a\times\mc W_b\times\mc S^n\times\mc Y^n$
%by the coding scheme $(f^{(a)},f^{(b)},g)$ and the memoryless MAC $Q$ coincides with that induced by the corresponding coding scheme $(\phi^{(a)}_t,\phi^{(b)}_t,\psi)$ and the finite-state MAC $P$.
%Hence, in this way, to any $(R,n,\eps)$-coding scheme on the memoryless MAC $Q$,
%it is possible to associate an $(R,n,\eps)$-coding scheme $(\phi^{(a)}_t,\phi^{(b)}_t,\psi)$ on the original finite-state MAC $P$.
%
Then, a standard arguments shows that the rate region
\begin{eqnarray}\label{regionequivMAC}
R_a &<& \mutinf(U_a;Z| U_b) \nonumber \\
R_b &<& \mutinf(U_b;Z| U_a) \nonumber \\
R_a+R_b&<& \mutinf(U;Z)\,,
\end{eqnarray}
is achievable on this MAC, where $U=(U_a,U_b)$ and $Z$ are random variables whose joint distribution factorizes as
\be P(U_a,U_b,Z)=\mu_a(U_a)\mu_b(U_b)Q(Z|U_a,U_b)\,,\label{factorizationbis}\ee
for some $\mu_a\in\mc P(\mc U_a)$, and $\mu_b\in\mc P(\mc U_b)$.
%For a positive integer $n$, let
%$f^{(n)}_a:\mc W_a^{(n)}\to\mc U_a^n$, and $f_b^{(n)}:\mc W_b^{(n)}\to\mc U_b^n$ be random encoders,
%with $|\mc W_a^{(n)}|=\lceil\exp(R_an)\rceil$, $|\mc W_b^{(n)}|=\lceil\exp(R_bn)\rceil$,
%and
%$$\left\{f^{(n)}_a(w_a),f^{(n)}_b(w_b):\,w_a\in\mc W_a^{(n)},w_b\in\mc W_b^{(n)}\right\}$$
%is a collection of independent r.v.s, with $f_i^{(n)}(w_i)$ taking values in $\mc U_i^n$ with product distribution $\mu_i{\otimes}\ldots\otimes\mu_i$, for each $i\in\{a,b\}$ and $w_i\in\mc W_i$. Then, it follows from the direct coding theorem for memoryless MACs \cite[Th.3.2, p. 272]{CsiszarKorner} that
%the average error probability of such a code ensemble converges to zero as $n$ grows large.
Now, one can restrict himself to choosing probability distributions $\mu_i$ in $\mc P(\mc U_i)=\mc P(\mc X_i^{\mc S_i})$ with the product structure
\be\label{product} \mu_i(u_i)=\prodd_{v_i\in\mc V_i}\pi_i(u_i(v_i)|v_i)\,,\quad u_i:\mc V_i\to\mc X_i\,,\ee
where $i\in\{a,b\}$, and $\pi$ is some memoryless stationary team policy, as in (\ref{policy}). Then, to any triple of r.v.s $(U_a,U_b,Z)$, with joint distribution as in (\ref{factorizationbis}), one can naturally associate  random variables $S$, $X_a:=U_a(q_a(S))$, $X_b:=U_b(q_b(S))$, and $Y$, whose joint probability distribution satisfies (\ref{factorization}). Moreover, it can be readily verified that
\be\label{equivMI}\ba{rcl}
\mutinf(X_a;Y| S, X_b) &=& \mutinf(U_a;Z| U_b) \\
\mutinf(X_b;Y| S, X_a) &=& \mutinf(U_b;Z| U_a)  \\
\mutinf(X;Y|S)&=& \mutinf(U;Z)\,.\ea\ee
Hence, if a rate pair $R=(R_a,R_b)$ belongs to the rate region $\mc R(\pi)$ associated to some memoryless stationary team policy $\pi$ (i.e.~if it satisfies (\ref{region})), then $R$ satisfies (\ref{regionequivMAC}) for the product probability distributions $\mu_a$, $\mu_b$ defined by (\ref{product}). This in turn implies that the rate pair is achievable on the original finite-state MAC $P$. The proof of achievability of the capacity region $\ov{\co}(\cup_{\pi}\mc R(\pi))$ then follows from a standard time-sharing principle (see, e.g., \cite[Lemma 2.2, p.272]{CsiszarKorner}).

\section{Converse to the coding theorem}
\label{sectconverse}
In this section, we shall prove that no rate outside $\ov{\co}(\cup_{\pi}\mc R(\pi))$ is achievable.
Lemma \ref{lemmaconverse1} shows that any achievable rate pair can be approximated by convex combinations of (conditional) mutual information terms.
For $\eps\in[0,1]$, define
\be\eta(\eps):=\frac{\eps}{1-\eps}\log|\mc Y|+\frac{\ent(\eps)}{1-\eps}\,,\label{etadef}\ee
and observe that
\be\lim_{\eps\to0}\eta(\eps)=0\,.\label{etalim}\ee
For every $t\ge1$, and $\mb\sigma\in\mc S^{t-1}$, define
\be\label{alphasdef}
\alpha_{\mb\sigma}:=\frac1n\P(S_{[t-1]}=\mb\sigma)\,.%\,,\qquad\mb\sigma\in\mc S^{(n)}\,,
\ee
Clearly, $\alpha_{\mb\sigma}\ge0$, and
\be
\!\summ_{\mb\sigma\in\mc S^{(n)}}\alpha_{\mb\sigma}=\frac1n\summ_{1\le t\le n}\summ_{\mb\sigma\in\mc S^{t-1}}\P(S_{[t-1]}=\mb\sigma)=1\,.\label{convex}\ee

\begin{lemma}\label{lemmaconverse1}
For a rate pair $R\in\R_+^2$, a block-length $n\ge1$, and a constant $\eps\in(0,1/2)$, assume that there exists a $(R,n,\eps)$-code.
Then,
\be
R_a+R_b\le\summ_{\mb\sigma\in\mc S^{(n)}}\alpha_{\mb\sigma}\mutinf(X_t;Y_t|S_t,S_{[t-1]}=\mb\sigma)+\eta(\eps)
\label{R1+R2}\ee
\be R_a\le
\summ_{\mb\sigma\in\mc S^{(n)}}\alpha_{\mb\sigma}\mutinf(X_t^{(a)};Y_t|X_t^{(b)},S_t,S_{[t-1]}=\mb\sigma)+\eta(\eps)\,.\label{R1}\ee
\be R_b\le
\summ_{\mb\sigma\in\mc S^{(n)}}\alpha_{\mb\sigma}\mutinf(X_t^{(b)};Y_t|X_t^{(a)},S_t,S_{[t-1]}=\mb\sigma)+\eta(\eps)\,.\label{R2}\ee
\end{lemma}
\proof
%Denote by ${\bf W}=(W_a,W_b)$ the message pair.
By Fano's inequality we have the following estimate of the residual uncertainty on the messages given the full decoder's observation
\be\ba{rcl}H(W|Y_{[n]};S_{[n]}) %&\le& %h(\eps) + \eps \log(|\mc W_1||\mc W_2|)\\
&\le&\ent(\eps)+\eps\log(|\mc W_a||\mc W_b|)\,.\ea\label{Fano1+2}\ee%%%%giacomo%%%%
For $1\le t\le n$, we consider the conditional mutual information term
$$\Delta_t:=\mutinf(W;Y_t,S_{t+1}|Y_{[t-1]},S_{[t]})\,,$$
and observe that
\be\ba{rcl}
\summ_{1\le t\le n}\Delta_t&=&
\ent(W|S_1)-\ent(W|S_{[n+1]},Y_{[n]})\\[5pt]
&=& \log(|\mc W_a||\mc W_b|)-\ent(W|S_{[n]},Y_{[n]})\,,\ea
\label{sumDeltat}\ee
since the initial state $S_1$ is independent of the message pair $W$, and the final state $S_{n+1}$ is conditionally independent of $W$ given $(S_{[n]},Y_{[n]})$.
On the other hand, using the conditional independence of $W$ from $S_{t+1}$ given $(S_{[t]},Y_{[t]})$, one gets
\be\ba{rcl}\Delta_t&=&\mutinf(W;Y_t,S_{t+1}|Y_{[t-1]},S_{[t]})\\[5pt]
&=&\mutinf(W;Y_t|Y_{[t-1]},S_{[t]})\\[5pt]
&=&\ent(Y_t|Y_{[t-1]},S_{[t]})-\ent(Y_t|W,Y_{[t-1]},S_{[t]})\\[5pt]
&\le&\ent(Y_t|S_{[t]})-\ent(Y_t|W,S_{[t]})\\[5pt]
&=&\mutinf(W;Y_t|S_{[t]})\,,
\ea\label{Deltat}\ee
where the above inequality follows from the fact that  $\ent(Y_t|Y_{[t-1]},S_{[t]})\le\ent(Y_t|S_{[t]})$,
since removing the conditioning does not decrease the entropy, while
$\ent(Y_t|W,Y_{[t-1]},S_{[t]})=\ent(Y_t|W,S_{[t]})$, as $Y_t$ is conditionally
independent from $Y_{[t-1]}$ given $(W,S_{[t]})$,
due to the absence of output feedback.
Since $(W,S_{[t]})-(X_t,S_t)-Y_t$ forms a Markov chain,  the data processing inequality
implies that
\be \mutinf(W;Y_t|S_{[t]}) \le \mutinf(X_t;Y_t|S_{[t]})%\le \mutinf(X_t;Y_t|S_{t})
\,.\label{IWY}\ee
%the former above inequality being implied by,
%%$$
%%\ba{rcl}
%%\mutinf(X_t;Y_t|S_{[t]})&=&\summ_s\P(S_t=s)\mutinf(X_t;Y_t|S_t=s)
%%%\summ_{s_{[t]},x,y}\P(S_{[t]}=s_{[t]})\P(X_t=x|S_{[t]}=s_{[t]})
%%%\log\frac{\P(Y_t=y|X_t=x,S_{[t]}=s_{[t]})}{\P(Y_t=y|S_{[t]}=s_{[t]})}
%%\ea
%%$$
%the latter inequality following from the concavity of the mutual information
%with respect to the input distribution (see e.g.~\cite[Lemma 3.5(d), p.50]{CsiszarKorner}).

%NOTE: Isn't this just the result that conditining reduces entropy?

%Now, for $i\in\{a,b\}$, and $1\le t\le n$, let us consider the $\mc P(\mc X_i)$-valued r.v.~$\Pi^{(i)}_t$, defined by
%$$\Pi^{(i)}_t(x):=\P(X^{(i)}_t=x|V^{(i)}_{t})\,.$$
%%$$\pi^2_t(v|V^2_{[t]}):=\P(V_t=v|\tilde {\mb S}^2_{[t]}\tilde{\mb\sigma}^2_{[t]})\,,$$
%%and
%%$$\Pi^{(i)}_t(u):=\P(X^{(i)}_t=u|V^{(i)}_t)\,.$$
%We shall also introduce the mutual information cost function
%$c:\mc S\times\mc P(\mc U_1)\times\mc P(\mc U_2)\to\R$
%defined by
%$$\ba{rcl} c(s,\pi_1,\pi_2)&:=\summ_{u_1,u_2,y}&\pi_1(u_1)\pi_2(u_2)P(y|s,u_1,u_2)\\
%&&\log\frac{P(y|s,u_1,u_2)}{\summ_{u_1,u_2}P(y|s,u_1,u_2)\pi_1(u_1)\pi_2(u_2)}\,.\ea$$
%%It is a standard fact that $c(s,\pi_1,\pi_2)$ is jointly concave in $(\pi_1,\pi_2)$.
%where the inequality above is implied by Jensen's inequality, the last line follows by the observation that there exists somw policy which only depends on conditioning on $S_t$.
By combining (\ref{Fano1+2}), (\ref{sumDeltat}), (\ref{Deltat}) and (\ref{IWY}), we then get
\be\ba{rcl}\!\!R_a\!+\!R_b\!\!&\!\!\le\!\!&\ds\frac1n\log(|\mc W_a||\mc W_b|)\\[5pt]
&\!\!\le\!\!&\ds\frac1{1-\eps}\frac1n\summ_{1\le t\le n}\mutinf(X_t;Y_t|S_{[t]})+\frac{\ent(\eps)}{n(1-\eps)}\\[5pt]
&\!\!\le\!\!&\ds\frac1n\summ_{1\le t\le n}\mutinf(X_t;Y_t|S_{[t]})+\eta(\eps)%\\
%&\!\!\le\!\!&\ds\frac1n\summ_{1\le t\le n}\mutinf(X_t;Y_t|S_{t})+\eta(\eps)\\
%&\!\!\!=\!\!&\ds\summ_{\mb\sigma\in\mc S^{(n)}}\alpha_{\mb\sigma}\mutinf(X_t;Y_t|s_t)+\eta(\eps)
\,.\ea\label{R1+R2preliminary}\ee
Moreover, observe that
$$\ba{rcl}\mutinf(X_t;Y_t|S_{[t]})&=&\ds\summ_{\mb\sigma\in\mc S^{t-1}}\P(S_{[t-1]}=\mb\sigma)\chi_{\mb\sigma}\\
&=&n\ds\summ_{\mb\sigma\in\mc S^{t-1}}\alpha_{\mb\sigma}\chi_{\mb\sigma}\,,\ea$$
where $\chi_{\mb\sigma}:=\mutinf(X_t;Y_t|S_t,S_{[t-1]}=\mb\sigma)$.
Substituting into (\ref{R1+R2preliminary}) yields (\ref{R1+R2}).

Analogously, let us focus on encoder $a$: by Fano's inequality, we have that
\be\label{Fano1}
\ent(W_a|Y_{[n]},S_{[n]})\le\ent(\eps)+\eps\log(|\mc W_a|)\,.\ee
For $t\ge1$, define
$$\Delta^{(a)}_t:=\mutinf(W_a;Y_t,S_{t+1}|W_b,Y_{[t-1]},S_{[t]})\,,$$
and observe that
\be\label{sumDelta1t}\ba{rcl}
 \!\!\summ_{1\le t\le n}\!\!\Delta^{(a)}_t\!\!
&\!\!=\!\!&\!\! \ent(W_a|S_1,W_b)-\ent(W_a|W_b,S_{[n+1]},Y_{[n]})\\[5pt]
&\!\!\ge\!\!&\!\! \log|\mc W_a|-\ent(W_a|S_{[n]},Y_{[n]})\,,
\ea\ee
where the last inequality follows from the independence between $W_a$, $S_1$, and $W_b$, and the fact that removing the conditioning does not decrease the entropy. Now, we have
\be\label{chain1}
\ba{rcl}
\Delta^{(a)}_t&\!\!=\!\!&\mutinf(W_a;Y_t,S_{t+1}|W_b,Y_{[t-1]},S_{[t]})\\[5pt]
&\!\!=\!\!&\mutinf(W_a;Y_t|W_b,Y_{[t-1]},S_{[t]})\\[5pt]
&\!\!=\!\!&\ent(Y_t|W_b,Y_{[t-1]},S_{[t]})-\ent(Y_t|W,Y_{[t-1]},S_{[t]})\\[5pt]
&\!\!\le\!\!&\ent(Y_t|W_b,S_{[t]})-\ent(Y_t|W,S_{[t]})\\[5pt]
&\!\!=\!\!&\mutinf(W_a;Y_t|W_b,S_{[t]})\,,
\ea
\ee
where the inequality above follows from the fact that $\ent(Y_t|W_b,Y_{[t-1]},S_{[t]})\le \ent(Y_t|W_b,S_{[t]})$
since removing the conditioning does not decrease the entropy,
and that $\ent(Y_t|W,Y_{[t-1]},S_{[t]})=\ent(Y_t|W,S_{[t]})$
due to absence of output feedback.
Observe that, since, conditioned on $W_b$ and $S_{[t]}$ (hence, on $X^{(b)}_t$),
$W_a-X_t^{(a)}-Y_t$ forms a Markov chain, the data processing inequality implies that
\be\label{dataproc1}\ba{rcl} \mutinf(W_a;Y_t|W_b,S_{[t]})&\le& \mutinf(X_t^{(a)};Y_t|X_t^{(b)},S_{[t]})\,.\ea\ee
%&\le&\mutinf(X_t^{(a)};Y_t|X_t^{(b)},S_{t})\,,\ea\ee
%the former inequality above being implied by the data processing inequality,
%the latter inequality following from the concavity of mutual information with respect to the input distribution.
By combining (\ref{Fano1}), (\ref{sumDelta1t}), (\ref{chain1}), and (\ref{dataproc1}), one gets
$$\ba{rcl}R_a\!\!&\!\!\le\!\!&\!\!\ds\frac1n\log|\mc W_a|\\[5pt]
&\!\!\le\!\!&\!\!\ds\frac1{n(1-\eps)}\summ_{1\le t\le n}\mutinf(X_t^{(a)};Y_t|X_t^{(b)}\!\!,S_{[t]})+\frac{\ent(\eps)}{n(1-\eps)}\\[5pt]
&\!\!\le\!\!&\!\!\ds\frac1n\summ_{1\le t\le n}\mutinf(X_t^{(a)};Y_t|X_t^{(b)},S_{[t]})+\eta(\eps)
\\[5pt]
%&\le&\ds\frac1n\summ_t\mutinf(X_t^{(a)};Y_t|X_t^{(b)},S_{t})+\eta(\eps)\\[5pt]
&\!\!=\!\!&\!\!\!\!\ds\summ_{\mb\sigma\in\mc S^{(n)}}\!\!\alpha_{\mb\sigma}\mutinf(X_t^{(a)};Y_t|X_t^{(b)},S_t,S_{[t-1]}=\mb\sigma)+\eta(\eps)
\,,\ea$$
%$\eta(\eps)$ and $\alpha_{\mb\sigma}$ being defined by (\ref{etadef}) and (\ref{alphasdef}), respectively.
%where the conditional mutual information function
%\mbox{$c_1:\mc S\times\mc P(\mc U^1)\times\mc P(\mc U^2)\to\R$ }
%is defined by
%$$\ba{rcl} c_1(s,\pi_1,\pi_2)&:=\summ_{u_1,u_2,y}\pi_1(u_1)\pi_2(u_2)P(y|s,u_1,u_2)\\
%&\log\frac{\summ_{u_2}P(y|s,u_1,u_2)\pi_2(u_2)}{\summ_{u_1,u_2}P(y|s,u_1,u_2)\pi_1(u_1)\pi_2(u_2)}\,.\ea$$
which proves (\ref{R1}).

In the same way, by reversing the roles of encoder $a$ and $b$, one obtains (\ref{R2}).
%where
%\mbox{$c_2:\mc S\times\mc P(\mc U^1)\times\mc P(\mc U^2)\to\R$ }
%is defined by
%$$\ba{rcl} c_2(s,\pi_1,\pi_2)&:=\summ_{u_1,u_2,y}\pi_1(u_1)\pi_2(u_2)P(y|s,u_1,u_2)\\
%&\log\frac{\summ_{u_1}P(y|s,u_1,u_2)\pi_1(u_1)}{\summ_{u_1,u_2}P(y|s,u_1,u_2)\pi_1(u_1)\pi_2(u_2)}\,.\ea$$
\qed

%Now, observe that, because of the mutual independence of $W_a$, $W_b$, and $S_{[t]}$, the r.v.s $X^{(a)}_t$ and $X^{(b)}_t$ are conditionally independent given $S_{[t]}$. Further
%We have the state sequence as: $\{S_t, t \geq 0\}$. The control action is a measurable mapping from feasible subset of $\sigma({\bf \tilde{s}}_{[t]})$ to ${\cal P}_{{\cal U}}$, hence control actions $P({\bf U}|{\bf \tilde{s}}_{[t-1]})$ take value in the product space of probability distributions on ${\cal U}$.
For $t\ge1$, let us fix some string $\mb\sigma\in\mc S^{t-1}$, and focus our attention on the conditional mutual information terms $\mutinf(X_t;Y_t|S_t,S_{[t-1]}\!\!=\!\!\mb\sigma)$, $\mutinf(X^{(a)}_t;Y_t|X^{(b)}_t\!,S_t,S_{[t-1]}\!\!=\!\!\mb\sigma)$, and $\mutinf(X^{(b)}_t;Y_t|X^{(a)}_t,\!S_t,S_{[t-1]}\!\!=\!\!\mb\sigma)$, appearing in the rightmost sides of (\ref{R1+R2}), (\ref{R1}), and (\ref{R2}), respectively. Clearly, the three of these quantities depend only on the joint conditional distribution of current channel state $S_t$, input $X_t$, and output $Y_t$, given the past state realization $S_{[t-1]}=\mb\sigma$. Hence, the key step now consists in showing that
\be\mb\nu_{\mb\sigma}(s,x,y):=\P(S_t=s,X_t=x,Y_t=y|S_{[t-1]}=\mb\sigma)\label{jointstateinputoutput}\ee factorizes as in (\ref{factorization}). This is proved in Lemma \ref{lemmaconverse2} below.

For $x_i\in\mc X_i$, $v_i\in\mc V_i$, and $\mb\sigma\in\mc S^{t-1}$,
let us consider the set $\Upsilon^{(i)}_{\mb\sigma}(x_i,v_i)\subseteq\mc W_i$,
$$
\Upsilon^{(i)}_{\mb\sigma}\!(x_i,v_i)\!:=\!\left\{\!w_i\!:\phi^{(i)}_t\!(w_i,q_i(\sigma_1),\ldots,q(\sigma_{t-1}),\!v_i)\!=\!x_i\!\right\}
$$
and the probability distribution $\pi^{(i)}_{\mb\sigma}(\,\cdot\,|v_i)\in\mc P(\mc X_i)$,
$$\pi^{(i)}_{\mb\sigma}(x_i|v_i):=\summ_{w_i\in\Upsilon^{(i)}_{\mb\sigma}(x_i,v_i)}|\mc W_i|^{-1}\,.$$

\begin{lemma}\label{lemmaconverse2} For every $1\le t\le n$, $\mb\sigma\in\mc S^{t-1}$, $s\in\mc S$, $x_a\in\mc X_a$, and $x_b\in\mc X_b$,
\be\label{asin}\mb\nu_{\mb\sigma}(s,x,y)=P(s)\pi^{(a)}_{\mb\sigma}(x_a|q_a(s))\pi^{(b)}_{\mb\sigma}(x_b|q_b(s))P(y|s,x)\,.\ee
\end{lemma}
\proof
First, observe that%, for any state $s\in\mc S$, input pair $x=(x_a,x_b)\in\mc X$, and output $y\in\mc Y$,
\be\label{P(S,X,Y|S-)}
\ba{rcl}
\mb\nu_{\mb\sigma}(s,x,y)\!\!\!&\!\!=\!\!
%&\P(S_t=s,X_t=x,Y_t=y|S_{[t-1]}=s_{[t-1]})\\
&\P(S_t=s|S_{[t-1]}=\mb\sigma)\nu_{\mb\sigma}(x|s)P(y|s,x)\\[5pt]
&=&P(s)\nu_{\mb\sigma}(x|s)P(y|s,x)
\ea
\ee
where $\nu_{\mb\sigma}(x|s):=\P(X_t=x|S_{[t]}=(\mb\sigma,s))$.
The former of the equalities in (\ref{P(S,X,Y|S-)}) follows from (\ref{channel}), while the latter is implied by the assumption that the channel state sequence is i.i.d..

Now, recall that, for $i\in\{a,b\}$, the current input satisfies $X^{(i)}_t=\phi^{(i)}_t(W_i,V^{(i)}_{[t]})$.
For $w\in\mc W$, let $\xi_w:=\P(X_t=x|S_{[t]}=(\mb\sigma,s),W=w)$. Then,
\be\label{P(X|S)}\ba{rcl}
\nu_{\mb\sigma}(x|s)&=&%\P(X_t=x|S_{[t]}=(\mb\sigma,s))\\&=&
\ds\summ\nolimits_w\xi_w
%\\[10pt]
%&&\qquad
\P(W=w|S_{[t]}=(\mb\sigma,s))\\[10pt]
&=&
\ds\sum\nolimits_w|\mc W_a|^{-1}|\mc W_b|^{-1}{\xi_w}\\[10pt]
&=&
\ds\!\!\!\sum_{w_a\in\Upsilon^{(a)}_{\mb\sigma}(x_a,q_a(s))}\!\!\!\!\!\!\!\!{|\mc W_a|}^{-1}
\summ_{w_b\in\Upsilon^{(b)}_{\mb\sigma}(x_b,q_b(s))}\!\!\!\!\!\!\!\!{|\mc W_b|}^{-1}\\[15pt]
&=&
\pi^{(a)}_{\mb\sigma}(x_a|q_a(s))\pi^{(b)}_{\mb\sigma}(x_b|q_b(s))\,,
\ea\ee
the second inequality above following from the mutual independence of $S_{[t]}$, $W_a$, and $W_b$.\
The claim now follows from (\ref{P(S,X,Y|S-)}) and (\ref{P(X|S)}).
\qed

Let us now fix an achievable rate pair $R=(R_a,R_b)$. By choosing $(R,n,\eps)$-codes for arbitrarily small $\eps>0$,
the inequalities (\ref{R1+R2}), (\ref{R1}), and (\ref{R2}), together with (\ref{etalim}) and (\ref{convex}), imply that $(R_a,R_b)$ can be approximated by convex combinations of rate pairs (indexed by $\mb\sigma\in\mc S^{(n)}$) satisfying (\ref{region}) for joint state-input-output distributions as in (\ref{jointstateinputoutput}). Hence, any achievable rate pair $R$ belongs to $\ov{\co}(\cup_{\pi}\mc R(\pi))$.

\textbf{Remark 1:}
For the validity of the arguments above, two critical steps were (\ref{P(S,X,Y|S-)}), where the hypothesis of i.i.d. channel state sequence has been used,
and (\ref{P(X|S)}), which only relies on the mutual independence of $W$ and $S_{[t]}$, this being a consequence of the assumption of absence of inter-symbol interference. In particular, the key point in (\ref{P(S,X,Y|S-)}) is the fact that the past state realization $\mb\sigma$ appears in $\nu_{\mb\sigma}$ only and not in $P(S_t)$. \hfill $\diamond$

\textbf{Remark 2: }
For the validity of the arguments above, it is critical that the receiver observes the channel state. More in general, it would suffice that the state information available at the decoder contains the one available at the two transmitters. In this way, the decoder does not need to estimate the coding policies used in a decentralized time-sharing.
\hfill $\diamond$

%\subsection{Information Stability and the Convergence of the Above Construction}
%
%The state process is i.i.d., and finite-state, hence is ergodic.

\section{Extensions to channels with memory and concluding Remarks}\label{sectconclusions}

The present paper has dealt with the problem of reliable transmission over finite-state multiple-access channels with asymmetric, partial channel state information at the encoders. A single-letter characterization of the capacity region has been provided in the special case when the channel state is a sequence of independent and identically distributed random variables.

It is worth commenting to which extent the results above can be generalized to channels with memory. Let us consider the case when the channel state sequence $\{S_t:\,t=1,2,\ldots\}$ forms a Markov chain with transition probabilities $\P(S_{t+1}=s_+|S_t=s)= P(s_+|s)$ which are stationary and satisfy the strongly mixing condition $P(s_+|s)>0$ for all $s,s_+\in\mc S$. Further, assume that there is no inter-symbol interference, i.e. $\{S_t:\,t=1,2,\ldots\}$ is independent from the message $W$, and that the state process is observed through quantized observations $V_t^{(i)}=q_i(S_t)$,  as discussed earlier.

For the generation of optimal policies in a multi-person optimization problem, whenever a dynamic programming recursion via the construction of a Markov Chain with a fixed state space is possible (see \cite{Yuksel} for a review of information structures in decentralized control), the optimization problem is computationally feasible and the problem is said to be {\em tractable}. In a large class of decentralized control problems, however, one faces intractable optimization problems. Let us elaborate on this further.

In team decision problems, one may assume that there is an \emph{a priori} agreement among the decentralized decision makers on who will do what, when the random variables take place. This approach is based on Witsenhausen's equivalent model for discrete stochastic control \cite{WitsenhausenEqui}, and makes the point that, indeed, all dynamic team problems are essentially static, with a much larger state space.

In the case of finite-state multiple-access channels with independent and identically distributed state sequences, by first showing that the past information is irrelevant, we observed that we could limit the memory space on which the dynamic optimization is performed. This is because, as observed in Remark 1, in the rightmost side of (\ref{P(S,X,Y|S-)}) the past state realization $\mb\sigma$ affects only the control $\nu_{\mb\sigma}(x|s)$, but not the current state distribution $P(S_t)$. In contrast, when the state sequence is a Markov chain, the past state realization $\mb\sigma$ does affect both the control $\nu_{\mb\sigma}(x|s)$ as well as the current state distribution $P(S_t|S_{[t-1]}=\mb\sigma)$. It is exactly such a joint dependence which prevents the proof presented here to be generalized to the Markov case.

Let us have a brief discussion for the case where there is only one transmitter. In this case,  the conditional probability distribution of the state given the observation history, $\Pi_t(\,\cdot\,):=\P(S_t=\,\cdot\,|V_{[t]})\in\mc P(\mc S)$,  can be shown to be a sufficient statistic, i.e.~the optimal coding policy can be shown to depend on it only. As a consequence, the optimization problem is tractable.
Such a setting was studied in \cite{YukTatISIT07}, where the following single-letter characterization was obtained for the capacity of finite-state single-user channels with quantized state observation at the transmitter and full state observation at the receiver:
$$C:=\int_{\mc P(\mc S)}\!\!\!\!\!\!\!\!\de\tilde P(\pi) \sup_{ P(X| \pi)\in\mc P(\mc X)}\left\{\!\summ_{s}   I (X;Y|s,\pi)\tilde P(s|\pi)\!\right\}$$
where %$\pi$ denotes the conditional distribution of the channel state given the quantized observation history, and
$\tilde P(s,\pi):=\tilde P(s|\pi)\tilde P({\pi})$ denotes the asymptotic joint distribution of the state $S_t$ and its estimate $\Pi_t$,
existence and uniqueness of which are ensured by the strong mixing condition.

For finite-state multiple-access channels with memory, a similar approach can successfully be undertaken only if the state observation is symmetric, namely if $q_a=q_b$. Indeed, in this case, the conditional state estimation $\Pi_t(\,\cdot\,)=\P(S_t=\,\cdot\,|V^{(a)}_{[t]})=\P(S_t=\,\cdot\,|V^{(b)}_{[t]})$ can be shown to be a sufficient statistic, and a single-letter characterization of the capacity region can be proved.

However, for the general case when the channel state sequence has memory and the state observation is asymmetric (i.e. $q_a\ne q_b$), the construction of a Markov chain (which would not incur a loss in performance) is not straightforward. The conditional measure on the channel state is no longer a sufficient statistic: In particular, if one adopts a team decision based approach, where there is a fictitious centralized decision maker, this decision maker should make decisions for all the possible memory realizations, that is the policy is to map the variables
$(W, V^{(a)}_{[t]}, V^{(b)}_{[t]})$ to $(X^{(a)}_t,X^{(b)}_t)$ decentrally, and the memory cannot be truncated, as every additional bit is essential in the construction of an equivalent Markov chain to which the Markov Decision Dynamic Program can be applied; both for the prediction on the channel state as well as the belief of the coders on each other's memory. Let us also elaborate a discussion in view of {\em common knowledge} of Aumann \cite{Aumann}: Information between two decision makers is common knowledge if it is measurable with respect to the sigma-fields generated by both of the local information variables at the decision makers. It is not usual in practical applications that all the local information is common knowledge. In such scenarios, one approach is to have the decision makers compute the conditional probability measures for the exogenous random variables and the actions of other decision makers for generating their optimal actions. For example, in the context of our problem in the paper, if we look for such {\em person-by-person optimal} policies, a policy of one of the encoders (say Encoder $a$) which uses the past will force the other encoder (Encoder $b$) to also use the past to second-guess the action of Encoder $a$, which requires the use of a policy with memory. Thus, adopting a person-by-person policy does not lead to useful structural results, in our context.

%Encoder $a$, in turn, will have to guess the guessing of Encoder $b$ with regard to its action, and this will lead to a %regress of belief on beliefs ad infinitum.

We instead adopted Witsenhausen's equivalent model to generate team policies, as also elaborated in \cite{Yuksel}, by having the encoders agree on which policies to adopt before random variables are realized. The approach in our paper showed that we can obtain a direct result when the channel state sequence is memoryless. However, when the channel state has memory, the past information provides useful information which is important for estimating the future channel states. As such, we cannot avoid the use of the information on the past channel state realizations. If one is to construct an equivalent state based on which coding policies are generated, the equivalent state needs to keep growing with time: The discussion in \cite{DasNarayan} provides such a block-level characterization and it seems we cannot go beyond this due to the non-tractability of the optimization problem. We note that if the encoders can exchange their past observations with a fixed delay, if they can exchange their observations periodically, or if they can exchange their beliefs at every time stage, then the optimization problem will be tractable.

One question of important practical interest is the following: If the channel transitions form a Markov chain, which is mixing fast, is it sufficient to use a finite memory construction for practical purposes? This is currently being investigated.

\begin{IEEEbiographynophoto}{Giacomo Como}
Giacomo Como is currently a Postdoctoral Associate at the Laboratory for Information and Decision Systems, Massachusetts Institute of Technology, Cambridge, MA. He received the BSc, MS and PhD degrees in Applied Mathematics from Politecnico di Torino in 2002, 2004 and 2008, respectively.  In 2006-07 he was Visiting Assistant in Research at the Department of Electrical Engineering, Yale University. His current research interests include the mathematics of information and control theory, network science, and coding theory.
\end{IEEEbiographynophoto}

\begin{IEEEbiographynophoto}{Serdar Y\"uksel}
Serdar Y\"uksel received his BSc degree in Electrical and Electronics Engineering from Bilkent University in 2001; MS and PhD degrees in Electrical and Computer Engineering from the University of Illinois at Urbana-Champaign in 2003 and 2006, respectively. He was a post-doctoral researcher at Yale University for a year before joining Queen's University as an assistant professor of Mathematics and Engineering at the Department of Mathematics and Statistics. His research interests are on stochastic and decentralized control, information theory and  applied probability. Dr. Y\"uksel serves on the IFAC (International Federation of Automatic
Control) Committee on Stochastic Systems.

\end{IEEEbiographynophoto}

\end{document}